\documentclass[runningheads]{llncs}
\usepackage{xcolor}
\usepackage{xspace}
\usepackage{url}
\usepackage[colorlinks=true, linkcolor=black, citecolor=red, urlcolor=black]{hyperref}

\usepackage{amsmath}
\usepackage{amssymb}	
\usepackage{pifont}

\usepackage{graphicx}
\usepackage{float}
\usepackage{wrapfig}

\usepackage{caption}
\usepackage{multirow}
\usepackage{rotating}
\usepackage{tabularx}
\usepackage[caption=false]{subfig}
\usepackage{booktabs} 

\usepackage{enumitem}

\usepackage{ifthen}
\usepackage{amssymb}
\usepackage{multirow}
\usepackage{soul}
\usepackage{listings}
\usepackage{color}

\usepackage{cite}
\usepackage{amsmath,amssymb,amsfonts}
\usepackage{textcomp}

\newboolean{showannotations}
\setboolean{showannotations}{true} 

\newboolean{shownames}
\setboolean{shownames}{true}

\newboolean{darkmode}
\setboolean{darkmode}{false} 

\newcommand{\newtextcolor}{blue}
\newcommand{\oldtextcolor}{red}
\newcommand{\assignementcolor}{orange}
\definecolor{highlightcolor}{rgb}{.99, 1, .0}
\sethlcolor{highlightcolor}
\definecolor{orcidlogocol}{HTML}{A6CE39}

\newboolean{showcomments}
\setboolean{showcomments}{true} 
\ifthenelse{\boolean{showcomments}}
{\newcommand{\nb}[2]{
    \fcolorbox{gray}{yellow}{\bfseries\sffamily\scriptsize#1}
    {$\blacktriangleright$#2$\blacktriangleleft$}
  }
  
}
{\newcommand{\nb}[2]{}
  
}

\definecolor{deepmagenta}{RGB}{204, 0, 204}

\newcommand{\eg}{e.g.,~}					
\newcommand{\ie}{i.e.,~}					
\newcommand{\Fig}[1]{Fig.~\ref{#1}} 		
\newcommand{\Sect}[1]{Section~\ref{#1}}     

\newcommand{\todo}[1]{
    \ifthenelse{\boolean{showannotations}}%
    {\ifthenelse{\equal{#1}{}}{\textcolor{red}{TODO}}{\textcolor{red}{TODO:~{#1}}}}%
    {}%
}

\newcommand{\assignedto}[1]{%
    \ifthenelse{\boolean{showannotations}}%
    {\textbf{\noindent\ding{46}\textcolor{white}{~\colorbox{\assignementcolor}{Assigned to:}}~\textcolor{\assignementcolor}{#1}\\}%
    }
    {}
}

\newcommand{\rem}[1]{%
    \ifthenelse{\boolean{showannotations}}%
    {\textcolor{\oldtextcolor}{\st{\textbf{#1}}}}%
    {}%
}

\newcommand\add[1]{%
    \ifthenelse{\boolean{showannotations}}%
    {\textcolor{\newtextcolor}{{\textbf{#1}}}}%
    {#1}%
}

\newcommand\rep[2]{%
    \ifthenelse{\boolean{showannotations}}%
    {\rem{#1}~\add{#2}}%
    {#2}%
}

\newcommand{\validity}[1]{\noindent {\bf {#1} validity.}}
\newcommand{\fakesection}[1]{\noindent {\bf {#1}.}}

\definecolor{darkgreen}{rgb}{0,0.6,0}
\definecolor{gray}{rgb}{0.5,0.5,0.5}
\definecolor{mauve}{rgb}{0.58,0,0.82}
\definecolor{gray}{rgb}{0.4,0.4,0.4}
\definecolor{darkblue}{rgb}{0.0,0.0,0.6}
\definecolor{lightblue}{rgb}{0.0,0.0,0.9}
\definecolor{cyan}{rgb}{0.0,0.6,0.6}
\definecolor{darkred}{rgb}{0.6,0.0,0.0}
\definecolor{armygreen}{rgb}{0.29, 0.33, 0.13}
\definecolor{antiquefuchsia}{rgb}{0.57, 0.36, 0.51}
\definecolor{coolblack}{rgb}{0.0, 0.18, 0.39}

\lstdefinelanguage{Relis}{
  morekeywords={PROJECT, CLASSIFICATION, SCREENING},
  morekeywords=[2]{Reviews, Conflict, on, resolved\_by, Criteria},
  keywordstyle=\color{mauve}\bfseries,
  keywordstyle=[2]\bfseries,
  morestring=[s][\color{darkred}]{"}{"},
  emph=[1]{DynamicList, Simple, string},
  emphstyle=[1]{\color{lightblue}},
}

\lstdefinelanguage{alloy}{
  keywords={%
    assert, pred, all, no, lone, one, some, check, run,
    but, let, implies, not, iff, in, and, or, set, sig, Int, int,
    if, then, else, exactly, disj, fact, fun, module, abstract,
    extends, open, none, univ, iden, seq,
    for, as, sum,
  },
  literate=%
  *{:}{{{\color[HTML]{2835C0}{$\colon$}}}}1
  {>}{{{\color[HTML]{2835C0}{>}}}}1
  {<}{{{\color[HTML]{2835C0}{<}}}}1
  {|}{{{\color[HTML]{2835C0}{|}}}}1
  {==}{{{\color[HTML]{2835C0}{$=$}}}}1
  {=}{{{\color[HTML]{2835C0}{$=$}}}}1
  {!=}{{{\color[HTML]{2835C0}{$\neq$}}}}1
  {&&}{{{\color[HTML]{2835C0}{$\land$}}}}1
  {||}{{{\color[HTML]{2835C0}{$\lor$}}}}1
  {<=}{{{\color[HTML]{2835C0}{$\le$}}}}1
  {>=}{{{\color[HTML]{2835C0}{$\ge$}}}}1
  {!in}{{{\color[HTML]{2835C0}{$\not\in$}}}}1
  {\\in}{{{\color[HTML]{2835C0}{$\in$}}}}1
  {=>}{{{\color[HTML]{2835C0}{$\implies$}}}}2
  {|=>}{{{\color[HTML]{2835C0}{$\Rightarrow$}}}}2
  {<=set}{{{\color[HTML]{2835C0}{$\subseteq$}}}}1
  {+set}{{{\color[HTML]{2835C0}{$\cup$}}}}1
  {*set}{{{\color[HTML]{2835C0}{$\cap$}}}}1
  {==>}{{{{\color[HTML]{2835C0}$\Longrightarrow$}}}}3
  {<==>}{$\Longleftrightarrow$}4
  {...}{$\ldots$}1
  {\\hl}{$\hline$}1
  {\\alpha}{$\alpha$}1
  {\\beta}{$\beta$}1
  {\\gamma}{$\gamma$}1
  {\\delta}{$\delta$}1
  {\\epsilon}{$\epsilon$}1
  {\\zeta}{$\zeta$}1
  {\\eta}{$\eta$}1
  {\\theta}{$\theta$}1
  {\\iota}{$\iota$}1
  {\\kappa}{$\kappa$}1
  {\\lambda}{$\lambda$}1
  {\\mu}{$\mu$}1
  {\\nu}{$\nu$}1
  {\\xi}{$\xi$}1
  {\\pi}{$\pi$}1
  {\\rho}{$\rho$}1
  {\\sigma}{$\sigma$}1
  {\\tau}{$\tau$}1
  {\\upsilon}{$\upsilon$}1
  {\\phi}{$\phi$}1
  {\\chi}{$\chi$}1
  {\\psi}{$\psi$}1
  {\\omega}{$\omega$}1
  {\\Gamma}{$\Gamma$}1
  {\\Delta}{$\Delta$}1
  {\\Theta}{$\Theta$}1
  {\\Lambda}{$\Lambda$}1
  {\\Xi}{$\Xi$}1
  {\\Pi}{$\Pi$}1
  {\\Sigma}{$\Sigma$}1
  {\\Upsilon}{$\Upsilon$}1
  {\\Phi}{$\Phi$}1
  {\\Psi}{$\Psi$}1
  {\\Omega}{$\Omega$}1
  {\\EOF}{\;}1
  ,
  sensitive=true,  
  morecomment=[l]//,%
  morecomment=[l]{--},%
  morecomment=[s]{/*}{*/},%
  morestring=[b]",
  commentstyle=\color[HTML]{00A108}\itshape,
  keywordstyle=\color[HTML]{2835C0}\bfseries,
  ndkeywordstyle=\bfseries,
}

\begin{document}
%
\title{How Users Employ Workarounds in Software Forms}

%
%
\author{MohammadAmin Zaheri \and Michalis Famelis \and Eugene Syriani}
\authorrunning{Zaheri et al.}
%
\institute{DIRO, Universit{\'e} de Montr{\'e}al \\
\email{\{zaherimo,famelis,syriani\}@iro.umontreal.ca}}
\maketitle              

\begin{abstract}
Workarounds enable users to achieve goals despite system limitations but expose design flaws, reduce productivity, risk compromising data quality, and cause inconsistencies. This study investigates how users employ workarounds when the data they want to enter does not align with software form constraints. Through a descriptive user study, we analyzed how workarounds originate and impact system design and data integrity. Understanding workarounds is essential for software designers to identify unmet user needs.

\end{abstract}

\section{Introduction}\label{sec:intro}

Software applications often require users to enter data for subsequent processing.
Examples include a citizen applying for benefits, a banking assistant recording transaction details in a financial system, a sales representative reporting customer visits in a CRM platform, a physician documenting patient symptoms and diagnostics in a medical application, a customer filing an insurance claim, or a photographer inputting parameters to optimize image rendering in a photo editor.
In most cases, these tasks are accomplished through forms, which serve as the primary interface for data entry.

In Information Systems, software applications often misalign with user needs
and goals~\cite{outmazgin2020workarounds}, leading them to improvise solutions.
These improvised {\em workarounds} are deliberate deviations from intended software usage that emerge when application forms and workflows fail to accommodate user needs~\cite{alter2014theory}. 
Sometimes, developers recognize workarounds as useful system features and integrate them in their designs.
For instance, developers designing a class diagram in Ecore within the Eclipse Modeling Framework often specify arbitrary values such as -1 or 1000 in the upper bound of an association to denote an unlimited number of targets, rather than using the UML-prescribed ``*''.
This workaround has become institutionalized as a documented feature in Eclipse because the cardinality text box only supports integer values.

The presence of workarounds can have positive and negative consequences~\cite{alter2015beneficial}.
Workarounds can reveal missing features, overlooked design flaws, or inefficient workflows.
For example, users might develop complex external processes to generate a report, unaware of an existing built-in functionality for the task.
While workarounds can be beneficial by enabling users to overcome system limitations, they also pose risks.
Deliberate deviations may lead to unintended consequences, such as data loss, system inconsistencies, or compromised data quality.
For instance, a user entering irrelevant data to bypass a required field might unintentionally
affect the accuracy of downstream reports, resulting in semantic inconsistencies where the system operates technically correctly but the data becomes unreliable~\cite{zaheri2024catch}.

Understanding the implications of workarounds is thus critical for improving
software design and usability. This is especially the case for low-code
applications, whose interfaces are typically forms and pages~\cite{sahay2020supporting}.
There is a lack of research on the workarounds employed by software
engineers~\cite{mubarkoot2023software}, and therefore by extension, citizen-developers~\cite{di2022low}
in low-code applications. We envision that insights into workarounds can be used 
to guide developers and product owners in identifying opportunities for new features, 
refine validation mechanisms, and enhance software maintenance.
For example, recognizing frequent use of comment boxes for unsupported data 
could lead to redesigning the form to better accommodate such information.

Here, we investigate how users handle misaligned cases 
where user goals or data requirements do not align with the constraints of software forms or workflows.
Through a descriptive analysis, we examine the origins of workarounds~\cite{soffer2023work} 
and their broader implications for system design and data integrity. 
The study provides insights into user behaviour when encountering misalignments.
We contribute a publicly available dataset of user workarounds 
and present key findings to advance research on workaround detection and user-centered software design.
Analyzing workarounds can inform future automatic detection mechanisms and recommendation systems for stakeholders in software design and usage.

We discuss related work in Section~\ref{sec:bg}, and outline the user study in
Section~\ref{sec:experiment}. We present our findings in
Section~\ref{sec:results}, and discuss them in Section~\ref{sec:discussion}. We
mention threats to validity in Section~\ref{sec:threats} and conclude in
Section~\ref{sec:conclusion}.

\section{Background and Related Work}\label{sec:bg}
%
Gasser defines workarounds in computing as follows~\cite{gasser1986integration}:
\textit{``working around means intentionally using computing in ways for which 
it was not designed or avoiding its use and relying on an alternative means of accomplishing work.''}
They emerge for various reasons, such as limitations in computing systems, challenges in existing workflows, or the need for quick solutions~\cite{alter2014theory}. 
Users often improvise when encountering technology misfits or data misalignments to complete their regular tasks.
In our context, a {\em technology misfit} is a mismatch between the technology in use and the practical needs or contingencies of everyday work. 
{\em Data misalignment} occurs when application processes or forms do not align with real-world data.
Workarounds are often seen as examples of bricolage and/or improvisation: 
{\em Bricolage} {``involves making do with available resources, occurring over short or long periods''}~\cite{levi1968savage};
{\em Improvisation} {``refers to bricolage in a short timeframe''}~\cite{miner2001organizational}.

Some workarounds are executed and disappear once the obstacle is overcome;
others become informally embedded into organizational
routines~\cite{white2023workarounds}.
Often, institutionalized workarounds become stepping stones for 
planned changes to processes and software~\cite{alter2014theory}.
%
%
Researchers have thus attempted to systematize and understand them.
A workaround ontology was developed by 
R{\"o}der et al.~\cite{roder2016toward},
while
Zainuddin et al.  proposed a taxonomy
~\cite{zainuddin2016developing}.

The data generated during the operation of information systems has been
extensively leveraged to detect workarounds, e.g., by mining logs
using deep learning methods~\cite{weinzierl2022detecting}, or for process
mining~\cite{beerepoot2022workarounds, van2022sword}. A common approach
is to track the sequence of actions and keystrokes users take when interacting
with software systems.
Workarounds can then be identified by comparing intended processes to actual user behaviour. 
Other studies focus on understanding the impact of workarounds, noting that users are often unaware of the potential consequences when they engage in these behaviours~\cite{drum2017workarounds, pernsteiner2018control}. 
Additionally, Schou et al. highlighted how unnoticed workarounds can have negative effects on organizational processes~\cite{schou2024we}.

Detecting workarounds in workflows often relies on interviews and case studies. 
Soffer et al. investigate the motivations for workarounds in processes and 
find that they arise from misalignments between user goals and official processes 
or conflicts among the goals themselves~\cite{soffer2023work}.
Van der Waal et al. investigated how workarounds evolve over time and whether they are adopted by individuals or groups, emphasizing the need for continuous monitoring~\cite{van2024emergence}. 
Davidson et al. explored the reasons behind workarounds, such as inadequacies in information systems, where users bypass formal processes simply to get their work done~\cite{davison2021coordination}. 
Wibisono studied how workarounds affect data quality, noting that users may interpret data in ways that introduce subjectivity and result in pseudo-data quality~\cite{wibisono2024workarounds}. 
Davidson et al. examined the role of no-code\slash low-code~\cite{di2022low} tools in enabling workarounds that deviate from established organizational processes~\cite{davison2024combining}.
Mubarkoot et al., taking the perspective of software compliance, highlighted
that while end-user workarounds have been studied, there is a lack of research on
workarounds for software engineers~\cite{mubarkoot2023software}.

\section{Experiment Design}\label{sec:experiment}

%
We aim to analyze the workarounds users do when interacting with
forms in the context of misalignment between data and form software interfaces.
We want to characterize their impact on user effort and data
quality, with respect to frequency, and systematic behaviour.
We pose the following research questions:
\begin{itemize}
  \item[\textbf{RQ1.}] What are the characteristics of workarounds users employ in software application forms?
  \begin{itemize}
    \item[\textit{RQ1.1.}]\textit{Which types of workarounds do users employ, if any?}
    \item[\textit{RQ1.2.}]\textit{How frequently do users employ workarounds ?}
    \item[\textit{RQ1.3.}]\textit{Which types of workarounds are most commonly associated with data misalignment types?}
  \end{itemize}
  We thus categorize workarounds, measure their frequency, and assess the quality of data input when users utilize workarounds.
  
  \item[\textbf{RQ2.}] \textit{Which form widgets are most commonly associated with workarounds?} \\
  We thus analyze patterns in the widget involved during workarounds.
  
  \item[\textbf{RQ3.}] How do users handle recurring data misalignments?
  \begin{itemize}
    \item[\textit{RQ3.1.}]\textit{How does the frequency of data misalignments affect user effort in workarounds?}
    \item[\textit{RQ3.2.}]\textit{To what extent do users develop systematic approaches when implementing workarounds?}
  \end{itemize}
  We therefore examine how repeated exposure to specific misalignments influences the time users spend on workarounds 
  and investigate whether these workarounds are applied systematically or tailored to individual cases.
\end{itemize}

\subsubsection{Setup}

We conducted an experiment~\cite{wohlin2012experimentation} 
to study user workarounds in software applications by analyzing interactions with 
forms for entering misaligned data. Participants used a Windows 11 PC 
(AMD Ryzen 7 5700U, 1.8GHz, 16GB RAM) with a 24-inch monitor and 
Python 3.11.7, and sessions were conducted via Microsoft Teams remote control.
%
Participants were tasked with entering prepared datasets into instrumented software applications we developed.
%
We extracted entered data, timing information, keystrokes and mouse clicks from
the generated logs to analyze the interactions with form widgets and application navigation.
We included cases where a participant entered and then corrected text.

\subsubsection{Experimental instruments}

We designed three simple applications with Python and CustomTkinter\footnote{\url{https://github.com/TomSchimansky/CustomTkinter}} 5.2.2:
a grading app ``TA'' for teaching assistants tasked with entering student grades for an exam;
a human resources app ``HR'' where HR professionals can complete an employee onboarding form;
and an event registration app ``REG'' for clerks to register participants to a conference.
All three consist of a main form that users can complete with records from the provided datasets.
They also allow 
navigating to previously entered records and modifying them, submitting feedback
in free-text globally for all the entered records, and reviewing all data before final submission.

We created three versions of TA: Open, Predetermined, and Controlled,
based on the different types of widgets used.
Lo et al. classified widgets as ``action'', ``data'', and ``static''~\cite{lo1996sizing}; 
Meneses et al. classified them as ``navigation'', ``input'', ``control'', and ``presentation''~\cite{meneses2014equivalence}. 
Our classification rather focuses on the affordance of widgets for user
workarounds, \ie
the degree of freedom that a widget offers to a user to manipulate. 
The Open version consists solely of widgets that allow users to
freely modify the input data they capture, 
such as text boxes and text areas with minimal or no validation rules.
The Predetermined version consists solely of widgets that contain preset data, 
requiring users to select from existing option, such as dropdown lists, combo boxes, 
radio buttons, checkboxes, and colour pickers.
The Controlled version consists solely of widgets that accept data
within a constrained (but potentially large) range of values
such as sliders, number spinners, file uploads, and date pickers.
The HR and REG apps offer a mix of all these three widget categories.

For each app, we synthesized a 
dataset consisting of 26 records for participants to enter.
Each dataset contains 11 records that can be easily entered in the application, 
\ie the data format and content align exactly with what the main form expects.
The remaining 15 records contain predefined data misalsignments.
We define six misalignment types that share the same intent for each application.
Since these types may overlap, we do not consider this separation as a partition but as a logical grouping of misalignments.
%
   \textbf{Explicit:} This misalignment is clearly visible before entering the record, 
  \eg in TA, the total is 105 (indicating that there was a bonus question), whereas normal grades are capped at 100. 
   \textbf{Implicit:} This misalignment is not immediately visible,
  \eg in TA, the record is marked to have completed the bonus, and the total grade is 95.
   \textbf{Excess:} The record contains more information than the form can capture,
  \eg the widget for the meal preference in REG
  cannot accommodate the detailed dietary requirements provided in the record. 
   \textbf{Nonstandard:} The data is formatted in an unconventional way,
  \eg a website rating widget in REG
  requires scores on a scale from 1 to 5, but the dataset contains the rating as
  a percentage.
   \textbf{Outlier:} The data provided is nonsensical,
  \eg in HR a record lists -1 hour for a widget capturing allowable remote work hours. 
   \textbf{Ambiguous:} The data can be interpreted in multiple ways,
  \eg it is not clear what multiple phone numbers within the same
  record mean and how to enter them in a single widget of HR.

Each dataset randomly includes 3 misalignment types with 5 instances of each type.
The 26 records of each dataset are randomly ordered for each participant.

\subsubsection{Experiment Process}
Each participant had an individual session where they
attempted to enter a provided dataset to the app.
Each session had three parts: introduction, tasks, and exit survey.
In the introduction, we presented the tasks and provided documentation about the apps.
The document included screenshots of completed forms to familiarize participants with the user interface and key features.
This part lasted 10 minutes.

We then asked participants to complete three data entry tasks, one with each app described above.
The version for the TA app was assigned randomly but balanced across all participants.
Participants were unaware of the goal of the experiment and were instructed only
to interact with three apps to enter data, with no mention of workarounds.
Although we did not enforce time limits for the experiment, participants took on average 1.5 hours to complete the experiment.
After completing the data entry tasks, participants completed a survey consisting of 12 questions structured as follows:
   \textbf{Background:} 1 open-ended question about the participant's primary occupation or field of study.
   \textbf{Ease of data entry:} 5 five-point Lickert-scale questions asking them to self-assess the ease and difficulty of entering the records.
   \textbf{Workarounds:} 5 five-point Lickert-scale questions on whether
  they resorted to workarounds for the misaligned records and on whether some workarounds were recurring.
   \textbf{Explanation and feedback:} 3 open-ended questions asking them to explain why some records were difficult to enter, to describe the workarounds they employed, and suggest how the forms could be adapted to handle the misalignments.

The experiment protocol, datasets, apps, exit survey, and statistical results are available in the replication package online.\footnote{
 \url{https://zenodo.org/records/14292103}
}

\subsubsection{Participants}

We used convenience sampling, sending invitations to research groups and communities within our network. 
No prior knowledge of any specific software was required. 
We hired 17 volunteer for the study: two in a pilot round to calibrate the experiment and refine the analysis, and 15 participants completed the experiment. 
Among them, we count 11 graduate students (including the two for the pilot), one bachelor student, and 5 professionals.
14 had a background in computer science, while 3 held degrees in other engineering disciplines.

\begin{figure}[t]
  \vspace{-1.5em}
  \centering
  \includegraphics[width=0.8\textwidth,height=6cm,keepaspectratio]{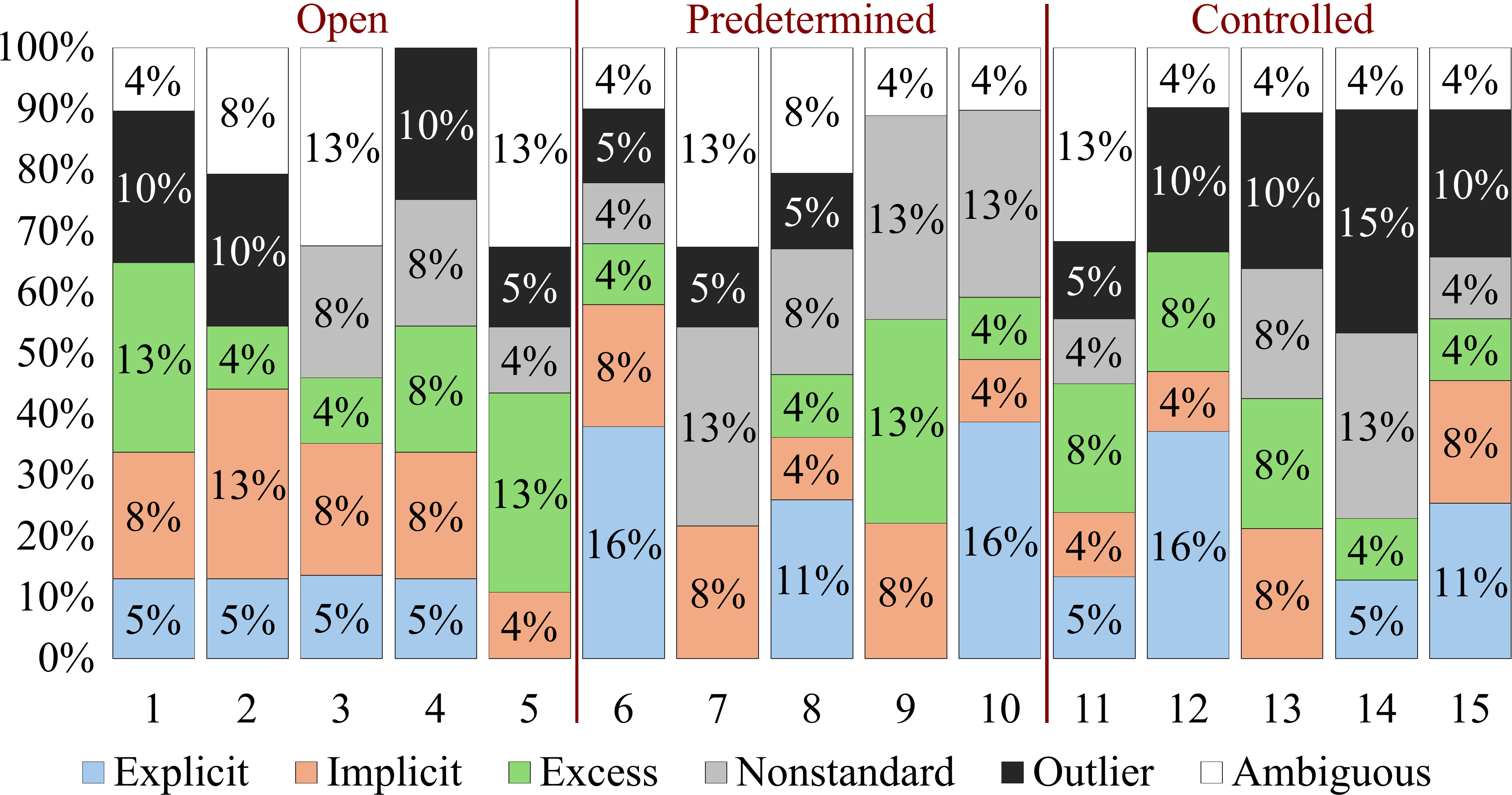}
  \caption{Distribution of the data misalignments across the 15 participants}
  \label{fig:assignments}
  \vspace{-1.5em}
\end{figure}
\Fig{fig:assignments} shows the distribution of the data misalignment types that
each participant faced across the 3 tasks.
Each participant encountered at least 4 misalignment types for a total 45 misaligned data.
We analyzed how participants employed workarounds for 120 Implicit, Excess, Ambiguous Nonstandard, 100 Outlier, and 95 Explicit  misalignments, for a total of 675 misaligned records.

\subsubsection{Metrics}\label{sec:metrics}

One author performed the initial labelling of the collected data, 
and another reviewed the labels. 
Conflicts were resolved collaboratively in a single session. 
We labelled each record that exposed a misalignment (\ie for which a workaround is expected) from each participant according to the workaround {\em type} and
{\em quality}.
We also collected the widget involved when a workaround occurs and the \textbf{widget category} (open, predetermined, or controlled).

{\em Workaround type}
reflects the ways users adapt to misalignments~\cite{zainuddin2016developing}.
{Data Adjustment (DA):} modifying the original data to align with application constraints; 
{Functional Adjustment (FA):} using the form's widgets or functionalities in unintended ways; 
{Fallback (FB):} resorting to an alternative, such as a general comment box, 
  to address the misalignment;
{Data Loss (DL):} truncating or omitting parts of the data, without any 
strategic approach;
{No Workaround (NW):} abandoning any attempt to resolve the misalignment. 

{\em Workaround quality}
assesses the integrity of the entered data and how much it differs from the intent of the provided misaligned data.
It estimates to what extent the original data and the entered data are semantically consistent~\cite{zaheri2024catch} and whether a human or machine can make similar inferences from them.
We give a score from 1 to 4 according to the following schema.
A quality score of 1 indicates significant loss of the provided misaligned data compared to what the participant entered. 
A score of 2 reflects partial loss of the misaligned data. 
A score of 3 means the participant retained the misaligned data indirectly, such as entering it elsewhere in the app or reporting it in a way that 
preserves the original intent.
A score of 4 is assigned when the participant manages to prominently input the misaligned data. 
If the misaligned data is completely skipped, it is considered as NW, with no quality score assigned.

From the save time of each record, we deduced the time each participant took to implement a workaround for each misaligned record.
Recall that, for each app, participants are provided with both normal and misaligned records.
Since each participant has a unique pace when entering data, regardless of data alignment, 
we calculate a distribution-sensitive central tendency measure as the baseline for normal record save times for each participant in each app.
To improve accuracy, we test the distribution of save times of each participant for each app using the Shapiro-Wilk test.
If the distribution is normal, the baseline is the mean.
For left-skewed or right-skewed distributions, the baseline is the 30th or 70th percentile respectively.
For heavy-tailed distributions (based on kurtosis), we compute the trimmed mean, excluding the lowest and highest 10\% of data points.
In all other cases, the baseline is the median.

{\em Relative workaround effort}
represents the time a participant takes to complete a workaround adjusted to the baseline.
It is computed for each misaligned record as the ratio of the save time when completing the workaround and the participant's baseline for that app.

{\em Systematicity score}
indicates to what extent participants consistently applied the same workaround type when encountering the same misalignment type within the same app.
For each unique misalignment type, we calculate the proportion of instances where the most common workaround type was used. 
Since each participant encountered three types of misalignment, with five instances of each per app, we averaged the scores across the three misalignment types 
to compute a single systematicity score for each participant per app.

\section{Results}\label{sec:results}

\subsection{Characteristics and frequency of workarounds (RQ1)}\label{sec:resultsrq1to3}

\begin{figure}[t]
    \vspace{-1.5em}
    \centering
    \includegraphics[width=.75\linewidth]{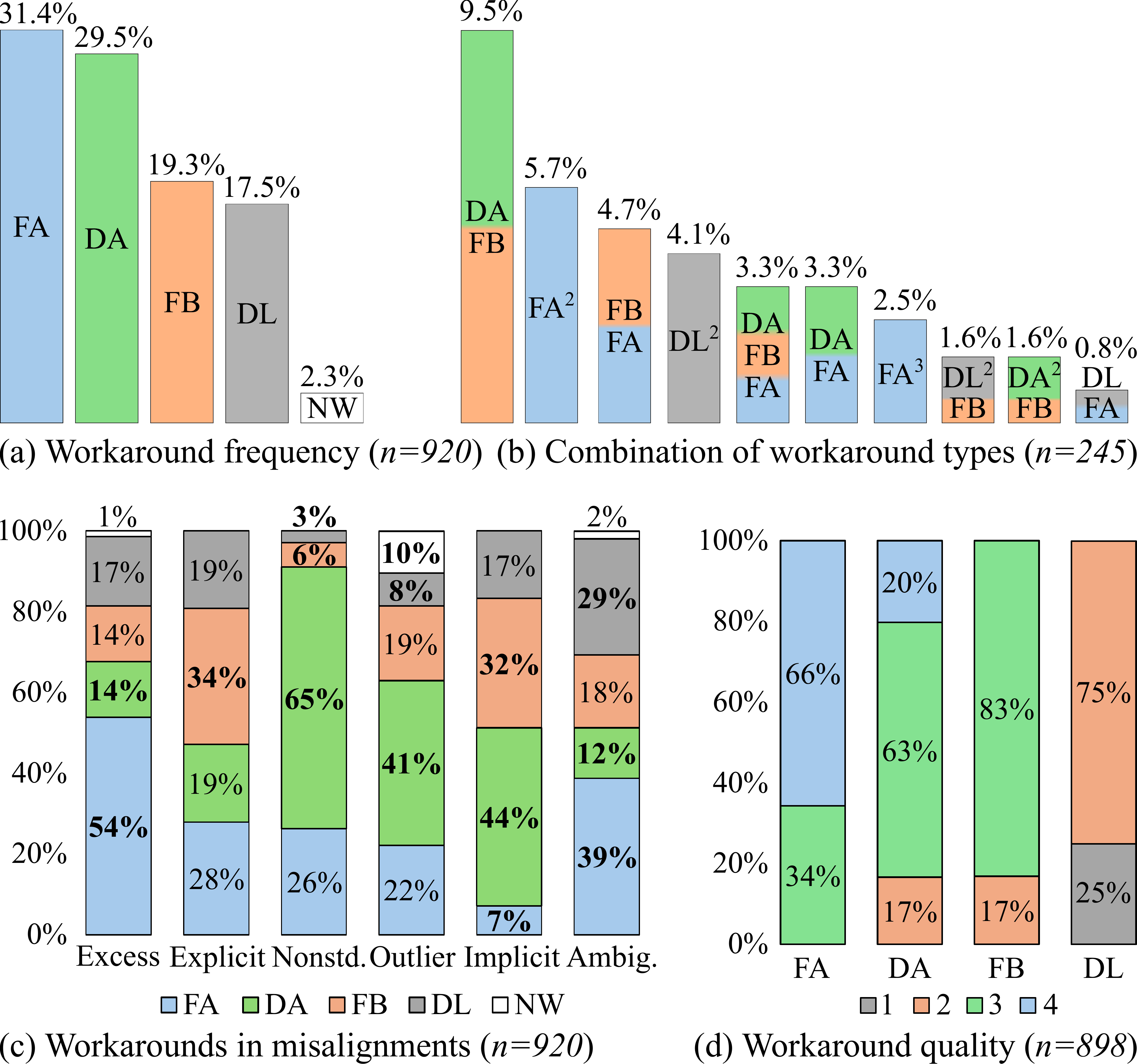}
    \caption{Characteristics of workaround types}
    \label{fig:wa_distrib}
    \vspace{-1.5em}
\end{figure}
Our analysis discovered cases where participants used 
multiple workaround types at once to address a single misalignment within the same record.
In these cases, we treated each workaround type separately, totaling of $920$ workarounds.
\Fig{fig:wa_distrib} presents the characteristics of workaround types for all these records.

As shown in \Fig{fig:wa_distrib}(a), participants almost always attempted workarounds when encountering a misalignment.
Five participants (22 records) ignored 
Outlier, Ambiguous, Excess, and  Nonstandard 
misalignments while using REG and open and controlled TA.
The most common workarounds are Functional and Data Adjustments, 
meaning that participants adjusted application functionalities 
(\eg establishing conventions or redefining widgets meaning) or manipulated 
the provided data to align with application constraints. 
In some cases, they employed Fallback methods, such as using a comment text box (\eg global feedback) to describe the issue.
Data Loss strategies, like truncation, were typically used as a last resort when no other workaround was feasible.

\Fig{fig:wa_distrib}(b) shows instances where multiple workaround types were combined. 
For the 245 records with more than one workaround type, 
the most frequent combination was Data Adjustment and Fallback.
This suggests that participants found the need to clarify the intent behind Data Adjustments to address subjective issues.
In some cases, participants combined Data and Functional Adjustments or multiple Functional Adjustments to achieve accurate data entry.
For example, in TA, participants repurposed the student feedback box and checkbox to tag records as suspicious, despite their original purpose.
For simpler scenarios, they truncated data for convenience or due to application constraints (\eg entering only the first out of three phone numbers when the widget allowed for only one). 
These limitations were occasionally documented using Fallback methods.
Fallback and Functional Adjustment are more versatile, frequently co-occurring with other workaround types.

\subsubsection{Workarounds and misalignments}

We conducted a Chi-Squared test to evaluate the association between workaround types and misalignment types.
The results indicated a statistically significant association $(\chi^2 = 295.11, df = 20, p < 0.000)$.
To measure the strength of this association, we calculated Cramer's V, which revealed a moderate effect size $(V = 0.283)$.
We conducted post hoc comparisons using pairwise Chi-Squared tests with Bonferroni correction $(p < 0.008)$.
\Fig{fig:wa_distrib}(c) presents the distribution of workaround types across misalignment types.
Significant associations are labelled in bold.
In the following, we describe what workarounds participants employed for each misalignment type. 

\textbf{Excess:}
participants primarily relied on Functional Adjustments 
to adapt widget functionalities for handling data beyond the app's native capabilities. 
Data Adjustments, though less frequent, were used to manipulate data to fit within the app's existing widgets.
\textbf{Explicit:} participants often resorted to Fallback methods to explain the Functional and Data Adjustments they applied to address the misalignment.
\textbf{Nonstandard:} participants primarily adjusted the data to normalize it for the app. 
Some attempts led to data loss, while others are complemented by a Fallback method to explain their approach.
\textbf{Outlier:} participants generally followed similar patterns as with Nonstandard misalignments.
However, they sometimes chose not to attempt any workarounds.
This suggests that, while participants may tolerate minor deviations from standard data, they are more likely to ignore data they perceive as nonsensical.
\textbf{Implicit:} participants primarily adjusted the data to perform minor modifications, such as correcting a date format, even if the data was already aligned with the app.
In some cases, they entered the data as-is but adjusted the functionality of a nearby widget to report the Implicit data, like repurposing the student feedback text box 
to report bonus points in TA.
%
\textbf{Ambiguous:} participants interpreted these misalignments in various ways, often choosing to either manipulate the data or adjust widget functionality within the app. 
This often led to data loss, as they retained only what they considered essential, such as keeping only one phone number in HR while discarding others they deemed unnecessary.

\subsubsection{Quality of workarounds}

\Fig{fig:wa_distrib}(d) shows the frequency of quality scores for each workaround type. 
Data Loss (DL) workarounds scored lowest, reflecting the simplicity of truncation at the cost of data integrity. 
This aligns with \Fig{fig:wa_distrib}(a), which shows participants' preference for workarounds that preserve data quality. 
Functional Adjustments (FA) and Data Adjustments (DA) yielded higher-quality
outcomes and were preferred over Fallback (FB) methods.
We observe an ordinality in the workaround types, 
where the quality of workarounds seems to improve progressively from 
DL to FA. 
Therefore, we pose the hypothesis, that there is an ordering between the workaround types with respect to the quality of the data entered, which can be summarized as: $\mathit{DL}<\mathit{FB}<\mathit{DA}<\mathit{FA}$.

To validate this hypothesis, we conducted a Spearman's rank-order correlation test 
for the relationship between workaround type (treated as ordinal) and workaround quality. 
The results showed a strong, significant positive correlation $(r_s=0.757, p<0.000, n=898)$.
This suggests that as participants transition from non-strategic data manipulation 
or simple Fallback methods (\eg descriptive comments) to employing data or Functional Adjustments, 
the quality of input data improves. 
Based on this and what we presented in \Fig{fig:wa_distrib}(b), 
Fallback methods are most effective when used alongside Functional or Data Adjustments. 
When used alone, they tend to produce lower-quality data.

    \textbf{Answer to RQ1: } When they encounter misalignments, users almost always employ a workaround.
    While workarounds result in varying levels of data quality, users generally choose approaches that maximize data integrity.
    They prioritize Functional and Data Adjustments, as these methods typically yield higher-quality outcomes. 
    In some cases, users opt for simpler alternatives, such as reporting issues in comment boxes or truncating data.

\subsection{Widget usage in workarounds (RQ2)}\label{sec:resultsrq4}

\begin{figure}[t]
  \vspace{-1.5em}
    \centering
    \includegraphics[width=0.75\linewidth]{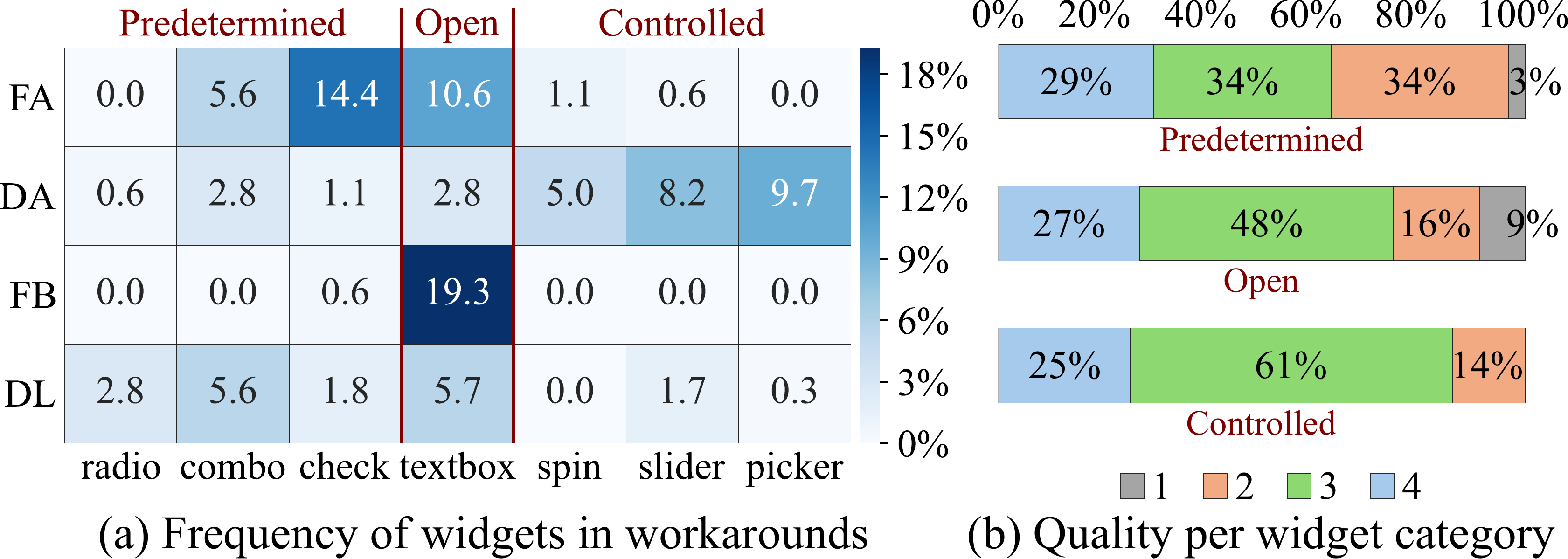}
    \caption{Widgets in workarounds ($n=898$)}
    \label{fig:widget_wa_qual}
    \vspace{-1.5em}
\end{figure}
As described in \Sect{sec:experiment}, there are three categories of app widgets
relative to how much user input freedom they offer: Open, Predetermined, and Controlled.
Across the $898$ records with workarounds (excluding the \textit{NW}), participants relied on text boxes for the Open widgets.
They attempted workarounds with radio buttons, combo boxes, and checkboxes for the Predetermined category.
Finally, they used number spinners, sliders, and date pickers for the Controlled category.

\subsubsection{Widgets involved in workarounds}

\Fig{fig:widget_wa_qual}(a) presents a heat map of widget usage across different types of workarounds.
Text boxes are most frequently used for Fallback workarounds (19.3\%), 
as participants often rely on Open widgets to input misaligned data in an unstructured manner. 
They also complement other workaround types, such as Functional Adjustment or Data Adjustment, 
by adding details or clarifying intentions. 
Occasionally, participants used text boxes for unintended purposes (\ie Functional Adjustment).
%
Sometimes, we observed Data Loss with Open text boxes, often stemming from participants' misinterpretation of the text box's intended purpose. 
For example, although HR provides a single text box capable of accommodating multiple phone numbers, participants tend to input only one number.
This behaviour occurs more frequently than attempting to adjust the data (\eg entering multiple numbers as a comma-separated value).

Among the Predetermined widgets used for workarounds, participants predominantly employed checkboxes for Functional Adjustments (14.4\%).
They used various combinations of checkboxes to represent specific concepts or data that could not be entered directly, 
or repurposed a checkbox to convey different meanings. 
For example, in HR, some participants selected ``fan'' as an item under employee amenities, even though the provided data had specified ``A/C''.
Unlike dropdown menus, combo boxes are editable and allowed some participants to store multiple values, 
such as multiple topics of interest in REG. 
Others assumed combo boxes accepted only one value, selecting a single option and causing Data Loss.
%
Radio buttons are less suitable for Functional Adjustments and lead to higher Data Loss, 
while checkboxes offer more flexibility by enabling multiple selections to approximate intended input.
%
Data Adjustment with radio buttons and Fallback use of checkboxes were rare (0.6\%), 
with the former involving selecting the closest option (\eg "Pasta (veg)" for "Pasta with meat" in REG) 
and the latter activating a text box for comments.

Data Adjustments were more common with Controlled widgets, 
especially date pickers, sliders, and number spinners. 
While these widgets enforce constraints, 
the results suggest that they they provide more flexibility than Predetermined widgets.
%
Participants creatively interpreted constraints, modified input formats, 
or manipulated data to work within widget restrictions while entering needed data. 
These adjustments were often complemented by Fallback entries in Open widgets.

\subsubsection{Impact of widgets workaround quality}

\Fig{fig:widget_wa_qual}(b) shows the quality of attempted workarounds across widget categories. 
We observed more frequently higher-quality workarounds (3--4) across all widget categories than lower-quality (1--2).
This suggests that participants preserved data integrity despite widget restrictions by employing creative solutions, such as repurposing fields to capture additional information.
For instance, they modified a combo box value to include job seniority in HR.
entered a bonus grade in an irrelevant text box in TA, 
and used the ``allergic'' checkbox, even though not marked in the data, 
with the meal preference checkbox in REG to indicate dietary preferences.

We observe a higher rate of data quality scores of 3 in Controlled widgets. 
This is because users often adhere to the constraints of these widgets and, 
when necessary, use complementary methods to capture data that cannot be directly entered due to the widget's controls. 
In contrast, Predetermined widgets show a higher rate of lower-quality data with a score of 2. 
This is expected, as Predetermined widgets impose more limitations 
and are less flexible for workarounds compared to Open and Controlled widgets.

Data with a score of 1, indicating significant loss in misalignment data, 
is more common in both Predetermined and Open categories. 
This often results from the single-widget effect discussed earlier, 
where users prefer to input a single value when the form provides only one widget 
(\eg a combo box or text box) that is conventionally meant to capture a single data item.

We conducted a Kruskal-Wallis H test to determine if there were statistically significant differences in workaround quality across the three widget categories.
Although it found a statistically significant difference among the categories ($H(2)=8.37, p= 0.015$), the post hoc pairwise comparisons using the Mann-Whitney U test with Bonferroni correction ($p < 0.017$) only reported significant differences between Predetermined and Controlled widget categories.
Therefore, the integrity of the data is better preserved when employing Controlled widgets in workarounds.

    \textbf{Answer to RQ2: } Users often rely on open widgets when employing Fallback methods.
    Their flexibility allows users to input unstructured or misaligned data, 
    supplement other workaround types, and even repurpose them for unintended applications. 
    Predetermined widgets are often used for Functional Adjustments but 
    are more prone to Data Loss due to their limited flexibility. 
    Controlled widgets, despite their constraints, 
    enable higher-quality data entry through creative user adaptations and complementary use of open widgets.

\subsection{Effort and Systematic Use of Workarounds (RQ3)}\label{sec:resultsrq5n6}

We now report on the user's behavior when encountering the same misalignment types multiple times.

\fakesection{Time effort}
We assess participant effort using the relative workaround effort (\Sect{sec:metrics}), 
measuring how much longer or shorter it took to save a misaligned record compared to baseline save time.
%
We used Spearman's test to examine the correlation between relative workaround effort and the order of repeated misalignment types.
%
Our results showed a significant weak negative correlation between the two variables $(r_s=-0.377, p=0.000, n=898)$.
%
This indicates that the more often participants encounter a specific type of misalignment, the less time they spend saving subsequent records of that type.
We can infer that participants recognize recurring \textit{patterns} and know how to handle the misalignment the more often they encounter it.

\fakesection{Systematicity}
To evaluate whether participants employed the same workarounds systematically when encountering the same misalignment type, 
we rely on the systematicity score (see \Sect{sec:metrics}).
Across all participants and apps ($n=45$), 88.1\% obtained a perfect score of 1, while the remaining case scored at least 0.87.
This indicates that participants consistently used the same workaround type for the same misalignment type.

\fakesection{Quality consistency}
To assess the consistency of workaround quality, we calculated the interquartile range (IQR) of workaround quality scores for each participant in each app. 
Our results showed that 90.5\% of the participants had an IQR$\le 1$. The maximum $IQR$ observed was 2 for one participant in one app.
The mean IQR is 0.74 indicating that participants maintained a consistent quality level when addressing different misalignments. 

\fakesection{Survey results}
The survey results further supported these findings.
80\% of participants agreed or strongly agreed with the statement, 
\textit{``I found another way to enter the records with difficulties (a workaround)''},
indicating they employed workarounds when encountering misalignments. 
Additionally, 93.3\% agreed or strongly agreed with, 
\textit{``I used the same workaround for records with similar difficulties''}, 
highlighting a systematic approach, as participants 
consistently applied the same type of workaround to similar misalignments.



    \textbf{Answer to RQ3: } As users encounter the same misalignment type more frequently, 
    they spend less time working around it. 
    This behaviour suggests that users develop systematic strategies and once 
    they successfully apply a workaround type for a specific misalignment, 
    they tend to reuse it for similar misalignments in the future 
    while maintaining a consistent level of quality.

\section{Discussion}\label{sec:discussion}
Our findings reveal distinct trends in how users handle misalignments in software application forms. 
Users strongly prefer workarounds over abandoning data entry, 
prioritizing strategies that preserve data integrity. 
Functional and Data Adjustments were the most frequent and effective workaround types, 
often resulting in higher data quality compared to Fallback or Data Loss. 
Users frequently combined multiple workaround types, 
particularly pairing Functional and Data Adjustments with Fallback methods to document additional details. 
While Fallback methods were less effective alone, 
they complemented other strategies by allowing users to capture otherwise unsupported information.

Users demonstrated a systematic approach when encountering repeated issues, becoming more efficient over time. 
A strong correlation between workaround type and data quality underscores their prioritization of accuracy. 
Open widgets like text boxes offered flexibility but were more prone to data loss due to users assuming one widget can only capture one piece of data. 
Predetermined widgets, like checkboxes, encouraged creative Functional Adjustments, 
while Controlled widgets, such as date pickers, facilitated higher data quality despite their 
constraints by guiding users toward more disciplined strategies. 
This challenges the assumption that more flexible widgets inherently result in better data capture, 
suggesting instead that well-defined constraints guide users toward disciplined, higher-quality strategies for workarounds.

These findings confirm the dual nature of workarounds. 
One the one hand, they expose design flaws, such as missing features or
misaligned workflows.
On the other hand, they pose risks to data integrity, especially when Fallback
or Data Loss strategies are used. 
Therefore, it is crucial that software designers proactively understand workarounds 
to identify unmet user needs and mitigate their risks through timely detection.
The results from our survey confirms that detecting workarounds and their reasons can uncover design problems. 
For instance, participants suggested features like \textit{``multi-option selectable for the dropdown''} 
or \textit{``more flexibility in attribute choices (topic of interest)''}, 
assuming combo boxes were limited to one piece of data, 
even though the designer intended the combo box to be editable, allowing users to add new values as needed. 
Early detection of such workarounds can address these issues. 
Another comment, \textit{``No need for manual input when Excel does the job''}, 
illustrates how designs unaligned with users' needs can drive them to adopt 
\textit{shadow systems} or \textit{complete bypass}~\cite{gasser1986integration}, 
supplementing or replacing the main system with alternatives.

\section{Threats to Validity}\label{sec:threats}

\validity{Construct}
One threat to validity is whether our workaround quality measurement accurately reflects true quality. 
While we based it on data retention to reduce subjectivity, it may not fully capture overall quality. 
Similarly, the systematicity score assumes that repeated use of the same workaround type indicates intentional behaviour, 
which may be influenced by factors like user background or problem interpretation. 

\validity{Internal}
Participants may have attempted more workarounds than usual due to the Hawthorne effect~\cite{adair1984hawthorne}.
To mitigate this, we carefully worded the invitation and instructions of the experiment to focus on data entry and software interaction, 
avoiding any mention of workarounds.
Additionally, while tasks and misalignments were randomized, 
earlier tasks in some sequences may have unintentionally helped participants perform better in later tasks.

\validity{External}
The tasks, application domains, and synthetic data used in our study may not fully represent real-world scenarios or capture the complexity of real-world misalignments, limiting the generalizability of our results.
While forms are pervasive in software and provide a suitable focus for this stage of research, 
the form-based scope may limit the applicability of our findings to other systems where workarounds may differ.
The participants' engineering and computer science background
may have influenced their approach to workarounds, affecting the generalizability of our findings. 
We tried to mitigate this by assigning them non-technical tasks, however it is
possible that, \eg computational thinking~\cite{wing2006computational} might influence the types of
workarounds employed.
Repeating the experiment with more diverse participants could help address this limitation.

\validity{Conclusion}
We had a small number of participants, which reduced the statistical power of our analysis. 
While we identified significant associations, the moderate strength of some statistical tests 
limits the generalizability of our findings to larger datasets and populations.
Finally, our interpretations of why participants chose certain workaround types or widgets may involve subjective judgment. 
To mitigate this, we first interpreted the results and then cross-referenced participants' responses from the exit survey to corroborate our findings.

\section{Conclusion}\label{sec:conclusion}
We investigated how users employ workarounds when software forms do not align with their data or goals. 
We collected a dataset of 920 user workarounds, and analyzed it for patterns in workaround types, their combinations, and their effects on system design and data quality.
%
%
We contribute insights into user behaviour when faced with misalignments, 
a publicly available dataset, 
and findings to support future research and guide designers in addressing user needs. 
Future work could explore non-form-based systems, 
include more diverse participants, 
and develop automated workaround detection to create more user-centered software.


\bibliographystyle{splncs04}
\bibliography{myref}

\begin{thebibliography}{10}
\providecommand{\url}[1]{\texttt{#1}}
\providecommand{\urlprefix}{URL }
\providecommand{\doi}[1]{https://doi.org/#1}

\bibitem{adair1984hawthorne}
Adair, J.G.: The hawthorne effect: a reconsideration of the methodological
  artifact. Journal of applied psychology  \textbf{69}(2), ~334 (1984)

\bibitem{alter2014theory}
Alter, S.: Theory of workarounds  (2014)

\bibitem{alter2015beneficial}
Alter, S.: Beneficial noncompliance and detrimental compliance: Expected paths
  to unintended consequences  (2015)

\bibitem{beerepoot2022workarounds}
Beerepoot, I.M.: Workarounds: The Path From Detection to Improvement. Ph.D.
  thesis, Utrecht University (2022)

\bibitem{davison2024combining}
Davison, R.M., Wong, L.H., Alter, S.: Combining low-code/no-code with
  noncompliant workarounds to overcome a corporate system’s limitations. MIS
  Quarterly Executive  \textbf{23}(3), ~5 (2024)

\bibitem{davison2021coordination}
Davison, R.M., Wong, L.H., Ou, C.X., Alter, S.: The coordination of
  workarounds: Insights from responses to misfits between local realities and a
  mandated global enterprise system. Information \& Management  \textbf{58}(8),
   103530 (2021)

\bibitem{di2022low}
Di~Ruscio, D., Kolovos, D., de~Lara, J., Pierantonio, A., Tisi, M., Wimmer, M.:
  Low-code development and model-driven engineering: Two sides of the same
  coin? Software and Systems Modeling  \textbf{21}(2),  437--446 (2022)

\bibitem{drum2017workarounds}
Drum, D., Pernsteiner, A., Revak, A.: Workarounds in an sap environment:
  impacts on accounting information quality. Journal of Accounting \&
  Organizational Change  \textbf{13}(1),  44--64 (2017)

\bibitem{gasser1986integration}
Gasser, L.: The integration of computing and routine work. ACM Transactions on
  Information Systems (TOIS)  \textbf{4}(3),  205--225 (1986)

\bibitem{levi1968savage}
Levi-Strauss, C.: The savage mind. Nature of Human Society, University of
  Chicago Press, Chicago, IL (Sep 1968)

\bibitem{lo1996sizing}
Lo, R., Webby, R., Jeffery, R.: Sizing and estimating the coding and unit
  testing effort for gui systems. In: Proceedings of the 3rd International
  Software Metrics Symposium. pp. 166--173. IEEE (1996)

\bibitem{meneses2014equivalence}
Meneses~Viveros, A., Hern{\'a}ndez~Rubio, E., V{\'a}zquez~Ceballos, D.E.:
  Equivalence of navigation widgets for mobile platforms. In: Design, User
  Experience, and Usability. User Experience Design for Diverse Interaction
  Platforms and Environments: Third International Conference, DUXU 2014, Held
  as Part of HCI International 2014, Heraklion, Crete, Greece, June 22-27,
  2014, Proceedings, Part II 3. pp. 269--278. Springer (2014)

\bibitem{miner2001organizational}
Miner, A.S., Bassof, P., Moorman, C.: Organizational improvisation and
  learning: A field study. Administrative science quarterly  \textbf{46}(2),
  304--337 (2001)

\bibitem{mubarkoot2023software}
Mubarkoot, M., Altmann, J., Rasti-Barzoki, M., Egger, B., Lee, H.: Software
  compliance requirements, factors, and policies: A systematic literature
  review. Computers \& Security  \textbf{124},  102985 (2023)

\bibitem{outmazgin2020workarounds}
Outmazgin, N., Soffer, P., Hadar, I.: Workarounds in business processes: a
  goal-based analysis. In: Advanced Information Systems Engineering: 32nd
  International Conference, CAiSE 2020, Grenoble, France, June 8--12, 2020,
  Proceedings 32. pp. 368--383. Springer (2020)

\bibitem{pernsteiner2018control}
Pernsteiner, A., Drum, D., Revak, A.: Control or chaos: impact of workarounds
  on internal controls. International Journal of Accounting \& Information
  Management  \textbf{26}(2),  230--244 (2018)

\bibitem{roder2016toward}
R{\"o}der, N., Wiesche, M., Schermann, M., Krcmar, H.: Toward an ontology of
  workarounds: A literature review on existing concepts. In: 2016 49th Hawaii
  international conference on system sciences (HICSS). pp. 5177--5186. IEEE
  (2016)

\bibitem{sahay2020supporting}
Sahay, A., Indamutsa, A., Di~Ruscio, D., Pierantonio, A.: Supporting the
  understanding and comparison of low-code development platforms. In: 2020 46th
  Euromicro Conference on Software Engineering and Advanced Applications
  (SEAA). pp. 171--178. IEEE (2020)

\bibitem{schou2024we}
Schou, P.K., Nesheim, T.: What we do in the shadows: How expert workers reclaim
  control in digitalized and centralized organizations through ‘stealth
  work’. Organization Studies  \textbf{45}(5),  719--744 (2024)

\bibitem{soffer2023work}
Soffer, P., Outmazgin, N., Hadar, I., Tzafrir, S.: Why work around the process?
  analyzing workarounds through the lens of the theory of planned behavior.
  Business \& Information Systems Engineering  \textbf{65}(4),  369--389 (2023)

\bibitem{van2022sword}
van~der Waal, W., Beerepoot, I., van~de Weerd, I., Reijers, H.A.: The sword is
  mightier than the interview: a framework for semi-automatic workaround
  detection. In: International conference on business process management. pp.
  91--106. Springer (2022)

\bibitem{van2024emergence}
van~der Waal, W., van~de Weerd, I., Beerepoot, I., Reijers, H.A.: The emergence
  and evolution of workarounds: A study of stability and change

\bibitem{weinzierl2022detecting}
Weinzierl, S., Wolf, V., Pauli, T., Beverungen, D., Matzner, M.: Detecting
  temporal workarounds in business processes--a deep-learning-based method for
  analysing event log data. Journal of Business Analytics  \textbf{5}(1),
  76--100 (2022)

\bibitem{white2023workarounds}
White, M.S.: Workarounds and shadow it--balancing innovation and risk. Business
  Information Review  \textbf{40}(3),  114--122 (2023)

\bibitem{wibisono2024workarounds}
Wibisono, A.: Workarounds produce pseudo-data quality: Insights from case
  studies. Procedia Computer Science  \textbf{234},  725--732 (2024)

\bibitem{wing2006computational}
Wing, J.M.: Computational thinking. Communications of the ACM  \textbf{49}(3),
  33--35 (2006)

\bibitem{wohlin2012experimentation}
Wohlin, C., Runeson, P., H{\"o}st, M., Ohlsson, M.C., Regnell, B., Wessl{\'e}n,
  A., et~al.: Experimentation in software engineering, vol.~236. Springer
  (2012)

\bibitem{zaheri2024catch}
Zaheri, M., Famelis, M., Syriani, E.: Catch me if you can: Detecting model-data
  inconsistencies in low-code applications

\bibitem{zainuddin2016developing}
Zainuddin, E., Staples, S.: Developing a shared taxonomy of workaround
  behaviors for the information systems field. In: 2016 49th Hawaii
  International Conference on System Sciences (HICSS). pp. 5278--5287. IEEE
  (2016)

\end{thebibliography}

\end{document}